\documentstyle[12pt,twoside]{article}
\newcommand{\bls}[1]{\renewcommand{\baselinestretch}{#1}}
\bls{1.8}
\def\xjoinrel{\mathrel{\mkern-4mu}}
\def\xlongrightarrow{\protect\@xlra}
\def\@xlra{\relbar\xjoinrel\relbar\xjoinrel\relbar\xjoinrel\relbar\xjoinrel\rightarrow}

\begin{document}

\vspace*{1cm}
\centerline{\large \bf On Zero-Mass Bound States in Super-Membrane Models}
\vspace{1cm}
\centerline{Jens Hoppe}
\vspace{1cm}
\centerline{Institut f\"ur Theoretische Physik}
\centerline{ETH-H\"onggerberg}
\centerline{CH-8093 Z\"urich}
\vspace{1cm}
\centerline{\large Abstract}
For the simplest case of a supermembrane matrix model, various symmetry
reductions are given, with the fermionic contribution(s) (to an
effective Schr\"odinger equation) corresponding to an attractive
$\delta$-function potential (towards zero-area configurations). The
differential equations are real, and are shown not to admit square-integrable
real solutions 
(even when allowing non-vanishing boundary conditions at infinity). Complex
solutions, however, are not excluded by this argument.

\vfil\eject  
Independent of the questions, whether supermembrane matrix models naturally
incorporate all topologies, as well as membrane interactions, it is important
to find out whether zero energy bound states exist in these models, or not.
I will address this question for the simplest such model, corresponding to a
supermembrane in \mbox{\raisebox{0.35ex}{$\scriptstyle |$}\hspace{-0.54em} \raisebox{0.45ex}{$\scriptstyle |$}\hspace{-0.1em}{\sf R}$^{1,3}$},
in a light-cone formulation and the two transverse
embedding coordinates being represented by 2$\times$2 matrices. The condition
for a zero energy state that it be annihilated by 2 hermitean supercharges
(whose square gives the Hamiltonian) leads to a single (though singular)
second order differential equation for one function of 6 variables.
Up to this point -- and slightly further (including the observation that
the desired wavefunction would have to have non-vanishing directions at $\infty$)
the analysis was carried out in [1]. Commenting on some subtleties in the
derivation (disappearence of the singularity when restricting to SU(2)-invariant
wavefunctions, resp.~conversion into boundary conditions, extra requirements
coming from the square-integrability of the full wavefunction -- including
fermions) I want to study in more detail non-trivial boundary conditions and
a further symmetry reduction to a differential equation in 2 independent
variables (whose bosonic part actually goes back more than 15 years [2]).
While the energy lowering effect of the fermions corresponds to an attractive
$\delta$-function towards zero area configurations, square integrable, real,
solutions nevertheless don't exist. The boundary conditions, however, arising
from the singularity at zero area, are shown to be complex, thus in general
coupling real and imaginary part of the wavefunction.

In the case of SU(2), and in slightly different notation, the supercharge $Q$
in [1] (eq.~4.29) reads:
\begin{equation}
Q = (\vec{\nabla}_1 - \mbox{i}\vec{\nabla}_2)\vec{D} + \mbox{i}(\vec{q}_1\times\vec{q}_2)\vec{\lambda}
\end{equation}
where $\vec{q}_1$ and $\vec{q}_2$ may be thought of as the position vectors of
two particles moving in \mbox{\raisebox{0.35ex}{$\scriptstyle |$}\hspace{-0.54em} \raisebox{0.45ex}{$\scriptstyle |$}\hspace{-0.1em}{\sf R}$^3$}, and the $\lambda_a$ ($a = 1,2,3$) are anticommuting
variables which together with $D^a := \displaystyle \frac{\partial}{\partial\lambda^a}$ are
satisfying $\{\lambda_a,\lambda_b\} = 0 = \{D^a,D^b\},\ \{D^a,\lambda_b\} = \delta_b^a$.
As the (square of the) Mass\mbox{(-operator)} is equal to the anticommutator of $Q$
with
$Q^{\dagger}\ (:= (-\vec{\nabla}_1 - \mbox{i}\vec{\nabla}_2)\vec{\lambda} - \mbox{i}(\vec{q}_1\times\vec{q}_2)\vec{D})$
zero-mass states $\Psi$ would have to be annihilated by $Q$ and $Q^{\dagger}$. \linebreak
The Ansatz [1]
\begin{equation}
\Psi = \psi + {\scriptstyle \frac{1}{2}}\epsilon^{abc}\psi_a\,\lambda_b\,\lambda_c
\end{equation}
leads to the equations
\begin{eqnarray}
(\vec{\nabla}_1 + \mbox{i}\vec{\nabla}_2)\psi & = & \mbox{i}\vec{q}\times\vec{\psi} \nonumber\\
\mbox{i}\,\vec{q}\,\psi & = & (\vec{\nabla}_1 - \mbox{i}\vec{\nabla}_2)\times\vec{\psi} \\
(\vec{\nabla}_1 + \mbox{i}\vec{\nabla}_2)\times\vec{\psi} & = & 0 \nonumber\\
\vec{q}\cdot\vec{\psi} & = & 0 \nonumber
\end{eqnarray}
where $\vec{q} := \vec{q}_1\times\vec{q}_2$. For $\vec{q} \not= 0$ they imply
\begin{equation} 
(-\vec{\nabla}^2 + q^2 + (\vec{\nabla}\,\ln q^2)\cdot\vec{\nabla} +
2\frac{\vec{q}\cdot{\vec{L}}}{q^2})\,\psi = 0 \quad\quad\quad ,
\end{equation}
as well as $\vec{q}_1\cdot{\vec{L}}=0=\vec{q}_2\cdot \vec{L}$, where
$\vec{L}:=-i\vec{q}_1 \times \vec{\nabla}_1 -i\vec{q}_2\times \vec{\nabla}_2$, 
$\vec{\nabla}$ denotes the gradient in \mbox{\raisebox{0.35ex}{$\scriptstyle |$}\hspace{-0.54em} \raisebox{0.45ex}{$\scriptstyle |$}\hspace{-0.1em}{\sf R}$^6$},
$q := \left| \vec{q}_1\times\vec{q}_2 \right|$ is the area spanned by the
2 particles (and the origin) in \mbox{\raisebox{0.35ex}{$\scriptstyle |$}\hspace{-0.54em} \raisebox{0.45ex}{$\scriptstyle |$}\hspace{-0.1em}{\sf R}$^3$},
and $\ln q^2$ is actually harmonic.
The simplicity of (4) could lead one to speculate that a similar
equation may well hold for higher dimensional gauge groups (SU(N)) and arbitrary
codimension $(d > 2)$, with $q$ being replaced by $\sqrt{V} = \sqrt{- \sum\limits_{i,j=1}^d Tr[X_i,X_j]^2}$,
or even for the original continuous membrane, where $\sqrt{V}$ is the area
of the actual surface embedded in $d+2$ dimensional space-time.

In any case, let us proceed with (4): though there exist more elegant
techniques of symmetry reduction, it may be instructive to do it step by step,
first letting $\psi$ depend only on
\begin{equation}
r_1 := \sqrt{\vec{q}_1^{\;2}}, \hspace{1em} r_2 := \sqrt{\vec{q}_2^{\;2}}, \hspace{1em} x := \cos <\hspace{-0.75em}\mbox{\raisebox{0.2ex}{\scriptsize )}}\;(\vec{q}_1, \vec{q}_2) \hspace{1cm}.
\end{equation}
(4) then becomes
\begin{equation}
\left( -\partial_1^2 - \partial_2^2 - (1 - x^2)\left(\frac{1}{r_1^2} + \frac{1}{r_2^2}\right)\partial_x^2 + r_1^2\,r_2^2(1 - x^2)\right)\,\psi = 0
\end{equation}
where the non-hermitean `fermionic' term
\begin{equation}
\vec{\nabla}\ln q^2\cdot \vec{\nabla}\psi =  \left(2\left(\frac{1}{r_1}\partial_1 + \frac{1}{r_2}\partial_2\right) - 2\left(\frac{1}{r_1^2} + \frac{1}{r_2^2}\right)x\,\partial_x\right)\,\psi
\end{equation}
has exactly cancelled some corresponding terms in the kinetic piece.
The finite part of the boundary ($r_1 = 0,\ r_2 = 0,\ x = \pm 1$) of the domain
on which (6) is defined precisely corresponds to $\vec{q} = 0$, where (3)
requires
\begin{equation}
(\vec{\nabla}_1 + \mbox{i}\vec{\nabla}_2)\psi = \vec{n}_1\!\left(\partial_1 - \frac{x}{r_1}\partial_x\right)\!\psi + \vec{n}_2\,\frac{1}{r_1}\,\partial_x\,\psi + \mbox{i}\vec{n}_2\!\left(\partial_2 - \frac{x}{r_2}\partial_x\right)\!\psi + \mbox{i}\,\vec{n}_1\,\frac{1}{r_2}\,\partial_x\,\psi = 0
\end{equation}
($\vec{n}_1$ and $\vec{n}_2$ are unit vectors into the direction of $\vec{q}_1$
and $\vec{q}_2$, respectively), as well as $(\vec{\nabla}_1 - \mbox{i}\vec{\nabla}_2)\times\vec{\psi} = 0$.
However, due to the fact that $\vec{\psi}$ has to be square-integrable, another important
condition must hold:
\begin{equation}
\int dq_1^{\,3}\,dq_2^{\,3} \left(\frac{\left|\vec{\nabla}\psi\right|^2}{q^2} + \frac{\mbox{i}}{q^2}\left(\vec{\nabla}_2\,\psi\,\vec{\nabla}_1\,\psi^{\ast} - \vec{\nabla}_1\,\psi\,\vec{\nabla}_2\,\psi^{\ast}\right)\right) < \infty \hspace{1cm}.
\end{equation}
Assuming, for the moment, $\psi$ to be real (in principle, complex boundary
conditions like (8) could couple the real and imaginary part of $\psi$),
(9) would imply
\begin{equation}
\int\limits_0^{\infty}dr_1 \int\limits_0^{\infty}dr_2 \int\limits_{-1}^{+1}dx\left(\frac{\psi_1^2 + \psi_2^2}{1 - x^2} + \left(\frac{1}{r_1^2} + \frac{1}{r_2^2}\right)\psi_x^2\right) < \infty \hspace{1cm},
\end{equation}
forcing $\psi$ to be constant on the boundary $r_1 = 0,\ r_2 = 0,\ x = \pm 1$.
I will be interested in the case of this constant being non-zero (otherwise,
(6) immediately implies $\psi \equiv 0$, as can be seen by dividing by
$(1-x^2)$, multiplying by $\psi^{\ast}$, and integrating).
Restricting now to wavefunctions that are also annihilated by
\begin{equation}
L = \left(\frac{r_1}{r_2} - \frac{r_2}{r_1}\right)\left(1 - x^2\right)\partial_x + x\left(r_1\,\partial_2 - r_2\,\partial_1\right)
\end{equation}
(the corresponding symmetry in the original variables is $\vec{q}_1 + \mbox{i}\vec{q}_2 \rightarrow \mbox{e}^{\mbox{\tiny i}\alpha}\left(\vec{q}_1 + \mbox{i}\vec{q}_2\right)$,
the only explicit remainder of Lorentz invariance),
3 reasonable choices of reduced variables are
\begin{equation}
r = \sqrt{r_1^2 + r_2^2}, \hspace{1em} y = \frac{2\,r_1\,r_2\,\sqrt{1 - x^2}}{r^2}
\end{equation}
(having assumed $\psi(-x) = \pm\psi(x)$, and taking, e.~g.,
$Z = r_1^2 - r_2^2$ as the variable on which $\psi$ should not depend) or [2]
\begin{eqnarray}
u & = & \frac{r}{\sqrt{2}}\sqrt{1 + \sqrt{1 - y^2}} =: r\cos\phi =: r\cos\frac{\theta}{4} \\
v & = & \frac{r}{\sqrt{2}}\sqrt{1 - \sqrt{1 - y^2}} =: r\sin\phi =: r\sin\frac{\theta}{4} \nonumber
\end{eqnarray}
in terms of which the conditions for a zero mass-state (apart from (9), resp.~(10))
read:
\begin{eqnarray}
\left(-\partial_r^2 - \frac{1}{r}\partial_r + \frac{4}{r^2}\left(y + \frac{1}{y}\right)\partial_y - \frac{4}{r^2}\left(1 - y^2\right)\partial_y^2 + \frac{1}{4}\,r^4\,y^2\right)\psi & = & 0 \\
\int\limits_0^{\infty}r^5\,dr\int\limits_0^1 ydy\left|\psi\right|^2 & < & \infty \nonumber
\end{eqnarray}
\begin{eqnarray}
\left(-\partial_u^2 - \partial_v^2 + u^2v^2 + \frac{u^2 + v^2}{u^2 - v^2}\left(\frac{1}{v}\partial_v - \frac{1}{u}\partial_u\right)\right)\psi & = & 0 \\
\int\limits_0^{\infty}du\int\limits_0^u dv\,u\,v\left(u^2 - v^2\right)\left|\psi\right|^2 & < & \infty \nonumber
\end{eqnarray}
\begin{eqnarray}
\left(-\partial_r^2 - \frac{1}{r}\partial_r - \frac{16}{r^2}\partial_{\theta}^2 + \frac{1}{8}\,r^4(1 - \cos\theta) + \frac{16}{r^2\sin\theta}\partial_{\theta}\right)\psi & = & 0 \\
\int\limits_0^{\infty}r^5\,dr\int\limits_0^{\pi}\sin\theta\,d\theta\left|\psi\right|^2 & < & \infty \hspace{0.2cm}.   \nonumber
\end{eqnarray}
Supposing $\left(\partial_x\psi\right)|_{x=0} = 0$ (even parity) one finds that
the normal derivative of $\psi(u,v)$ (resp.~$\partial_{\theta}\psi$) should
vanish at $u = v$, resp.~$\theta = \pi$ (this could be used to extend the
differential operator in (15) to the $v > u \ge 0$ octant, with $\psi$ then
vanishing on the half-lines $v = 0$ and $u = 0$, while imposing $\psi(u,v) = \psi(v,u)$).
In any case, let me mention various forms that eq.~(15), $H\psi$ = 0, takes
when conjugating $H$ by a function, i.~e.~considering $\sqrt{\rho}\,H\,\frac{1}{\sqrt{\rho}}$ for
various $\rho$. In all cases,
\begin{equation}
\int\left(\frac{u}{v} - \frac{v}{u}\right)\left(\psi_u^2 + \psi_v^2\right) < \infty
\end{equation}
accordingly transformed, provides important additional information on the
wavefunction(s).
Defining $\widetilde{\psi} = \sqrt{uv(u^2-v^2)}\,\psi$, (15)
can be written as 
\begin{eqnarray}
\left(\left(\partial_u + \frac{1}{2u} + \frac{u}{u^2-v^2}\right)\left(-\partial_u + \frac{1}{2u} + \frac{u}{u^2-v^2}\right)\right. \nonumber \\
+ \left(\partial_v + \frac{1}{2v} - \frac{v}{u^2-v^2}\right)\left(-\partial_v + \frac{1}{2v} - \frac{v}{u^2-v^2}\right)    \nonumber\\
\left. +\ u^2\,v^2 + \frac{2}{u}\partial_u + \frac{2}{v}\partial_v - \frac{1}{u^2} - \frac{1}{v^2}\right)\widetilde{\psi} & = & 0 \\
\int\limits_0^{\infty}du\int\limits_0^u dv\,\left|\psi\right|^2 & < & \infty \hspace{0.5cm}, \nonumber
\end{eqnarray}
with $\widetilde{\psi}$ now vanishing on the boundary (as long as $\psi$ was
finite there). Multiplying (18) by $\widetilde{\psi}$ (if real), and integrating, one
gets positive contributions from the first 3 terms, $+ \int\widetilde{\psi}\left(U^{\dagger}U + V^{\dagger}V + u^2\,v^2\right)\widetilde{\psi}$,
while getting $-\int\limits_0^{\infty}du\left.\left(\frac{1}{v}\widetilde{\psi}^2\right)\right|_{v=0}$
from the remaining 4 (which represent $\vec{\nabla}\ln q^2\,\vec{\nabla}\,\psi$).
Non-zero constant boundary conditions for $\psi(v=0)$ (resp. $\widetilde{\psi} \sim u\sqrt{uv}$)
will make the negative integral diverge, implying that at least one of the
positive integrals must also diverge. A less obvious change of measure,
$\widehat{\psi} := \sqrt{\frac{u^2-v^2}{uv}}\,\psi$, leads to
$\left(U^{\dagger}U + V^{\dagger}V + u^2\,v^2 + \left(\frac{1}{u^2} + \frac{1}{v^2}\right)\right)\widehat{\psi} = 0$,
resp.
\begin{eqnarray}
\left(-\partial_u^2 - \partial_v^2 + u^2\,v^2 + \frac{3}{4}\left(\frac{1}{u^2} + \frac{1}{v^2}\right) - \frac{u^2+v^2}{\left(u^2-v^2\right)^2}\right)\widehat{\psi} & = & 0 \\
\int u^2\,v^2\,|\widehat{\psi}|^2 & < & \infty \hspace{0.5cm}, \nonumber
\end{eqnarray}
while in the variables q=uv, p=${\rm \frac{u^2-v^2}{2}}$, one gets
\begin{eqnarray}
(-\partial_p^2 - \frac{1}{p}\partial_p - \partial_q^2 + \frac{1}{q} \partial_q
+ \frac{q^2}{2 \sqrt{p^2 + q^2} } ) \psi =0,  \nonumber \\
\int \frac{dpdq}{\sqrt{p^2 + q^2}} pq |\psi |^2 < \infty , \\
\int \frac{p}{q} ( \psi_p ^2 + \psi_q^2 ) dpdq < \infty. \nonumber
\end{eqnarray}

Let me finish by pointing out that the normalization condition, $\int
r_1^2r_2^2 |\psi|^2 dr_1dr_2<\infty$, together 
with (10) and $\psi=1$ on $x=\pm 1$ is not possible, as (10) implies $\psi_x
\to 0$ for fixed $r_1$, $x$ close to 1, and $r_2 \to \infty$, while
$r_2^2|\psi|^2\to 0$ (for $
r_2 \to \infty$) implies $\psi \to 0$, hence a
contradiction if $\psi_x \to 0$ and $\psi=1$ on $x=1$.

{\bf Acknowledgement} 
I am grateful to the members of our institute, as well as M.~Bordemann,
F.~Finster and M.~Struwe, and, in particular, J. Fr\"ohlich for
valuable discussions. 

{\bf References}
\begin{tabbing}
[1] \hspace{1em} \= B.~de Wit, J.~Hoppe, H.~Nicolai; Nucl.~Phys. {\bf B\,305} (1988) 545. \\[1ex]
[2]              \> J.~Goldstone, J.~Hoppe; unpublished notes (1980). \\
\end{tabbing}

\end{document}